\documentstyle[11pt,preprint,amssymb,amsbsy,eqsecnum,aps,epsf]{revtex}

\renewcommand{\vec}{\boldsymbol}

\begin{document}
\everymath{\displaystyle}
\draft
\title{
Deuteron spin oscilations and rotation as a universal method of the N-N
scattering amplitude study. }
\author{V. G. Baryshevsky, K. G. Batrakov and S. Cherkas}
\address{
Institute of Nuclear Problems 220050, Minsk, Republic of Belarus}
\date{\today}
\maketitle

\begin{abstract}
We consider the effects of the deuteron spin
rotation and oscillations
at the
matter which is proportional to the
real part of the
deuteron spin-dependent
forward scattering amplitude.
That gives a possibility of a
direct
measurement of this quantity.
Spin-dependent forward scattering amplitude of
the deuteron on an unpolarized  proton is discussed in terms of
the Glauber multiscattering theory. This amplitude is
determined by the nucleon rescattering, nonsphericity of the deuteron and
spin-dependent nucleon-nucleon scattering amplitude.
Thus deuteron spin oscillation phenomenon represents a methosd for
N-N scattering amplitude investigation, including its real part, over
a broad energy range.
\end{abstract}

\narrowtext

\section{Introduction}

The most fundamental
principles of our understanding of particle physics
are
analyticity and unitarity.
Through analyticity we can get dispersion relation between the real
and imaginary parts of the forward scattering amplitude.  Experimental
checking of dispersion relations is of great importance.  While
the imaginary part of zero angle amplitude is expressed through the
total cross section using unitarity relation the measurement of the
real part presents difficulties.  The traditional method of
measurement of the real part of nuclear scattering amplitude
utilizes interference with the presumed known Coulomb amplitude
dominating at small scattering angles \cite{nik,akch}.  For measuring the
elastic
scattering cross section over such a small range of the momentum
transferred the special methods  based on the
spectrometry of the recoil nucleons are applied. It is need to avoid
rescattering by target for the accurate measurement of
the momentum and the angle of the recoil nucleon .The low density
targets \cite{nik}
(supersonic hydrogen jet )
or special detectors \cite{akch}  are used for this aim.

It has been shown in \cite{bary1,bary2} that there is an
unambiguous method which makes direct measurement of the
real part of the spin dependent forward scattering
amplitude possible.
This technique  based on the phenomena
of particle beam spin rotation
and
oscillation in a matter  \cite{bary1,bary2}
uses  measurement of spin rotational angle
under the condition of a transmission experiment ---
the so-called spin rotation experiment.

For the particles with spin
$S\ge 1$  the
spin oscillation and rotation exist
even in unpolarized targets and its value  does not decrease
with particle energy grows \cite{bary1,bary2,omega}
at high energies.

Spin rotation and oscillation can be described by the particle
spin-dependent refractive index of
a medium which is proportional to the forward scattering amplitude by a target
particle. For particles     with
spin $S\geq 1$ ($\Omega ^{\hbox {-}}$-hyperon, deuteron)
forward scattering amplitude by an
unpolarized nucleon has the form:
\begin{equation}
\hat{F}(0)=T_{0}+T_{2}({\bf Sn)^{2}~~,}  \label{f0}
\end{equation}
where ${\bf S}$ is the particle spin operator and ${\bf n}$ is the
direction of the particle momentum.
Analysis by the frame of the Glauber multiscattering theory
\cite{bary2,omega}
shows   that $T_{2}$ is determined  by  rescattering of
colliding particle constituents only i.e. the first term of Glauber series
(single scattering) does not contribute to  $T_{2}$. So, one has the unique
possibility to observe nucleon rescattering in deuteron-proton collision and
constituent quarks rescattering in $\Omega ^{\hbox {-}}$-hyperon proton
collision \cite{bary2,omega}.

There exist two different contributions to the $T_{2}$ term. The first one is
due to the nonsphericity of the $S\geq 1$ particle and the second one
is due to the spin-dependent elastic scattering amplitude of the constituents.
At present work $T_{2}$ is calculated  for deuteron-proton collision and
it is found that the
contribution due to spin dependence of N-N scattering amplitude decreases
with  energy increasing and at the deuteron energy
$\gtrsim 5~Gev$ only the contribution due to deuteron nonsphericity survives.

Deuteron spin oscillation phenomenon is proportional to the product
of a real to imaginary parts of the N-N scattering amplitude
and increases with the energy growth at high energies.
So a
new possibility for checking dispersion relations arises. In
particular, at high and super high energies we may check
derivative dispersion relations for a scattering amplitude.

\section{ The rotation and oscillation phenomenon of a deuteron spin}

The motion of a particle with spin  inside the matter  can be
described by the refractive index   \cite{bary1}
\begin{equation}
\hat{n}=1+\frac{2\pi \rho }{k^{2}}\hat{F}(0)~.  \label{ind_ref1}
\end{equation}
Here $\rho $ is the density of scatterers in the matter (the number of
scatterers in $1~cm^{3}$ ), $k$ is the wave number of an incident particle. $%
\hat{F}(0)$ is the zero-angle elastic scattering amplitude which is an
operator in spin space of the incident particle. The dependence of the
amplitude $\hat{F}(0)$ on the orientation of colliding particle spins gives
rise to quasioptic effects (spin rotation, spin oscillation and dichroism)
under the passage of a particle through the medium.

For particles with spins $S\geq 1$ the spin rotation  arises even in
passing through unpolarized targets \cite{bary1,bary2,omega}.

Consider the propagation of deuteron through the unpolarized medium in
details. Let the spin state of a deuteron being incident on a target be
characterized by an initial spin wave function $\psi _{0}$. Then the spin
wave function in the target can be written as:
\begin{equation}
{\psi}(z)=\exp \{ik\hat{n}z\}{\psi}_{0}~.  \label{psi}
\end{equation}
If the target is unpolarized, the zero angle scattering amplitude is
determined by the deuteron spin properties only (\ref{f0}). In this case
$T_{0}=F_{0}(0)~~$, $T_{2}=F_{1}(0)-F_{0}(0)~$, where
$F_{1}(0)$
and $F_{0}(0)$ correspond to zero - angle scattering amplitudes for
the deuteron with
spin projection $|S_{z}|=\pm 1$ and $ S_{z}=0$, ${\bf n}$
is the direction of the incident deuteron momentum.

Equation (\ref{f0}) contains the terms only with square-law spin
dependence (we consider T - invariant interactions, therefore
the odd powers of spin in the amplitude should be absent). Taking the
quantization z - axis parallel to ${\bf n}$ and denoting the magnetic
quantum number through $m$ we obtain from equation (\ref{ind_ref1})
and (\ref{f0}):
\begin{equation}
n_{m}=1+\frac{2\pi \rho }{k^{2}}F_{m}(0),\ \ \ n_{m}=n_{m}^{\prime
}+in_{m}^{\prime \prime }~,  \label{refraction}
\end{equation}
where $n_{m}^{\prime }$ ,$n_{m}^{\prime \prime }$ are the real and imaginary
parts of the refractive indices for the particle in an \ eigenstate with spin
operator projection $S_{z}=m$.

It follows from (\ref{f0}), that the states with quantum numbers $m$
and $ -m$ are described by the same refraction indexes, however
$n_{\pm 1}\neq n_{0}$. This difference determines such effects as
spin rotation and oscillation.

In general case the deuteron spin wave function at a medium entry can be
written as:
\begin{equation}
{\psi}_{0}=\{{\mathfrak a}\,e^{i\delta _{-1}},{\mathfrak b}%
\,e^{i\,\delta _{0}},{\mathfrak c}\,\,e^{i\,\delta _{1}}\}.  \label{psi0}
\end{equation}
The polarization properties of the particle with spin $S=1$ are expressed
through spin vector ${\bf S}$ and rank $2$ tensor
$\hat{Q}_{ij}=3/2(\hat{S}_{i}%
\hat{S}_{j}+\hat{S}_{j}\hat{S}_{i}-4/3\delta _{ij})$ \cite{dau}. Using
equations (\ref{ind_ref1}),(\ref{psi}),(\ref{psi0}) it is possible to find
their evolution as a function of a particle way length inside the target. In
particular, for the polarization vector
($<{\bf S>=}\frac{<{\psi}{\bf | \hat S|}
{\psi}>}{\mid\psi\mid^2}$) we have the following
expressions:

\widetext
\begin{eqnarray}
&<&{S}_{x}>=\sqrt{2}\,{e^{-(n_{0}^{\prime \prime }+\,n_{1}^{\prime
\prime })z}}\,{\mathfrak b}(\,{\mathfrak a}\cos [\delta _{-1}-\delta
_{0}+(n_{1}^{\prime }-n_{0}^{\prime })\,z]+  \nonumber \\
&&\,{\mathfrak c}\,\cos [\delta _{0}-\delta _{1}+(n_{0}^{\prime
}-n_{1}^{\prime })\,z])/|\psi |^{2}  \nonumber \\
&<&{S}_{y}>=\sqrt{2}\,{e^{-(n_{0}^{\prime \prime }+\,n_{1}^{\prime
\prime })z}}\,{\mathfrak b}\,(\,{\mathfrak a}\sin [\delta _{-1}-\delta
_{0}+(n_{1}^{\prime }-n_{0}^{\prime })\,z]  \label{spin} \\
&&+\,{\mathfrak c}\,\sin [\delta _{0}-\delta _{1}+(n_{0}^{\prime
}-n_{1}^{\prime })\,z])/|\psi |^{2}  \nonumber \\
&<&{S}_{z}>={{e^{-2\,z\,n^{\prime \prime }_1}}}\left(
-{{{\mathfrak a}
^{2}}}+{\mathfrak c}{{^{2}}}\right) /|\psi |^{2}
\nonumber \end{eqnarray}

\narrowtext

\bigskip

And for the tensor rank $2$:

\widetext
\begin{eqnarray}
&<&{Q}_{xx}>=\biggl\{-[{\mathfrak a}^{2}+{\mathfrak c}%
^{2}]\,e^{-2n_{1}^{\prime \prime }\,z}/2+{\mathfrak b}^{2}\,e^{-2\,n_{0}^{%
\prime \prime }\,z}+  \nonumber \\
&&3\,{e^{-2n_{1}^{\prime \prime }\,z}}\,{\mathfrak ac}\cos [\delta
_{-1}-\delta _{1}]\biggr\}/|\psi |^{2}  \nonumber \\
&<&{Q}_{yy}>=\biggl\{-[{\mathfrak a}^{2}+{\mathfrak c}%
^{2}]\,e^{-2\,n_{1}^{\prime \prime }\,z}/2+{\mathfrak b}^{2}\,e^{-2\,n_{0}^{%
\prime \prime }\,z}-  \nonumber \\
&&3\,{e^{-2n_{1}^{\prime \prime }\,z}}\,{\mathfrak a}\,{\mathfrak c}\cos
[\delta _{-1}-\delta _{1}]\biggr\}/|\psi |^{2}  \nonumber \\
&<&{Q}_{zz}>=\biggl\{[{\mathfrak a}^{2}+{\mathfrak c}^{2}]\,e^{-2%
\,n_{1}^{\prime \prime }\,z}-2\,{\mathfrak b}^{2}\,e^{-2\,n_{0}^{\prime
\prime }\,z}\biggr\}/|\psi |^{2}  \label{quadr} \\
&<&{Q}_{xy}>=3\,{e^{-2n_{1}^{\prime \prime }z}}\,{\mathfrak a}\,{%
\mathfrak c}\,\sin [\delta _{-1}-\delta _{1}]/|\psi |^{2}  \nonumber \\
&<&{Q}_{xz}>=\biggl\{\frac{3}{\sqrt{2}}{e^{-(n_{0}^{\prime \prime
}+\,n_{1}^{\prime \prime })z}}\,{\mathfrak b}(-{\mathfrak a}\cos [\delta
_{-1}-\delta _{0}+(n_{1}^{\prime }-n_{0}^{\prime })z]  \nonumber \\
&&+{\mathfrak c}\cos [\delta _{0}-\delta _{1}+(n_{0}^{\prime }-n_{1}^{\prime
})z])\biggr\}/|\psi |^{2}  \nonumber \\
&<&{Q}_{yz}>=\biggl\{\frac{3}{\sqrt{2}}{e^{-(n_{0}^{\prime \prime
}+\,n_{1}^{\prime \prime })z}}\,{\mathfrak b}(-{\mathfrak a}\,\sin [\delta
_{-1}-\delta _{0}+(n_{1}^{\prime }-n_{0}^{\prime })z]  \nonumber \\
&&+{\mathfrak c}\,\sin [\delta _{0}-\delta _{1}+(n_{0}^{\prime
}-n_{1}^{\prime })z])\biggr\}/|\psi |^{2}~~,  \nonumber
\end{eqnarray}

\narrowtext

where $|\psi |^{2}=[{\mathfrak a}^{2}\,+{\mathfrak c}^{2}]e^{-2\,n_{1}^{%
\prime \prime }\,z}+{\mathfrak b}^{2}\,e^{-2\,n_{0}^{\prime \prime }\,z}$
and $z$ is the particle way length inside the medium. From these expressions
it follows that in general case the spin dynamics of ~deuteron is
characterized by a superposition of two rotations in clockwise and
counter-clockwise directions. Let's consider some particular cases. If the
initial polarization vector is normal to the particle momentum so, the
initial populations and phases of the states with the quantum numbers $m$
and $-m$ are equal  the
components $<S_{y}>$ and $<S_{z}>$  remain zero during the whole time of
particle penetration through the medium and $<S_{x}>$ oscillates. The
components of quadrupole tensor of deuteron oscillate too. If the initial
polarization vector is directed at an acute angle to the momentum direction
the polarization vector motion looks like rotation in
{fig.\ref{rotation}.  If the initial polarization vector is
directed at an obtuse angle to the momentum direction the spin
rotates in the opposite direction.

\section{ Eikonal approximation for the spin particles.}

\bigskip\
The phenomena of spin oscillation and rotation are defined by the value of
$T_{2}$.The two factors give contribution to $T_{2}$ : 1) nonsphericity
of a deuteron;
2) spin dependence of nucleon - nucleon scattering amplitude.
In (\cite{bary2},\cite{omega})  the  eikonal  Glauber approximation \cite{gl,sit}
was used for the study of oscillation and rotation phenomena.
 The spin
dependence of scattering amplitude isn't taken into account in the
traditional Glauber theory.   Therefore, only the contribution of
1) was studied. Now, for the account of 1) and 2) we should  use the spin
eikonal Glauber approximation.  Consider proton-deuteron elastic
scattering.  According to \cite{Wat} the scattering operator for
two colliding particles is written as
\renewcommand{\Lambda}{{\mathcal T}} \begin{equation} \Lambda
=V_{int}+V_{int}\Omega V_{int}  \label{basic} \end{equation} \bigskip
where $\Omega =\frac{1}{E-H+i0}$, $H=K+V_{int}$ is Hamiltonian of the
system, $V_{int}$ is the operator of deuteron - proton interaction,
$K$ is the noninteracting Hamiltonian part for the
$d-p$ system. Making the
transformations $\frac{1}{E-H+i0}=\frac{1}{E-K+i0}\left( 1+V_{int}\frac{1}{
E-H+i0}\right) $ we derive $\Lambda =V_{int}+V_{int}\Omega _{0}\Lambda $,
where$\ \ \ \Omega _{0}=\frac{1}{E-K+i0}$ is the free propagator function of
''deuteron + proton'' system. The interacttion part of Hamiltonian is
written as  $V_{int}=V_{1}+V_{2}$ where $V_{1}$, $V_{2}$ are
''proton - proton'' and
''neutron - proton'' interaction operators. By analogy we can write:

\begin{equation}
\Lambda _{1}=V_{1}+V_{1}\Omega _{01}\Lambda _{1}\ , \ \ \ \ \ \
\Lambda _{2}=V_{2}+V_{2}\Omega _{02}\Lambda _{2}~,  \label{objelem}
\end{equation}

Where $\Omega _{01}$, $\Omega _{02}$ are the free propagators of the
''proton + proton'' and ''neutron + proton''systems.

Expressing $V_{1}$ and $V_{2}$ through $\Lambda _{1}$ and $\Lambda _{2}$
from (\ref{objelem}): $V_{1}=\Lambda _{1}(1+\Omega _{01}\Lambda _{1})^{-1}$,
$V_{2}=\Lambda _{2}(1+\Omega _{02}\Lambda _{2})^{-1}$ and substituting
results in (\ref{basic}) we obtain \widetext
\begin{eqnarray}
\Lambda &=&(1-V_{int}\Omega _{0})^{-1}V_{int}=\left\{ 1-\left( \Lambda
_{1}(1+\Omega _{01}\Lambda _{1})^{-1}+\Lambda _{2}(1+\Omega _{02}\Lambda
_{2})^{-1}\right) \Omega _{0}\right\} ^{-1}  \nonumber \\
&&\times \bigl( \Lambda _{1}(1+\Omega _{01}\Lambda _{1})^{-1}+ \Lambda
_{2}(1+\Omega _{02}\Lambda _{2})^{-1}\bigr) \approx\left\{ 1+\Lambda
_{1}\Omega _{0}+\Lambda _{2}\Omega _{0}\right\} \{ \Lambda _{1}-\Lambda
_{1}\Omega _{01}\Lambda _{1}+\Lambda _{2}
 \nonumber \\
&&-\Lambda _{2}\Omega _{02}\Lambda
_{2}\}
\approx \Lambda _{1}+\Lambda _{2}+\Lambda _{1}(\Omega _{0}-\Omega
_{01})\Lambda _{1}+ \Lambda _{2}(\Omega _{0}-\Omega _{02})\Lambda
_{2}+\Lambda _{1}\Omega _{0}\Lambda _{2}+\Lambda _{2}\Omega _{0}\Lambda _{1}
\label{multiple}
\end{eqnarray}
\narrowtext
\medskip
Considering  proton scattering by resting deuteron
in the impulse approximation for the proton kinetic energy much
greater than the bound energy of the deuteron we can neglect the difference
between $\Omega _{0}$ and $\Omega _{0i}$:

\begin{equation}
\Lambda \approx \Lambda _{1}+\Lambda _{2}+\Lambda _{1}\Omega _{0}\Lambda
_{2}+\Lambda _{2}\Omega _{0}\Lambda _{1}~.  \label{approx}
\end{equation}
It can be written for small angle
scattering:
\begin{equation}
<out|\Lambda |in>=-\delta ^{(3)}({\bf p}_{out}-{\bf p}_{in})\frac{p}{(2\pi
)^{2}\varepsilon (p)}F({\bf q,n})~,  \label{def_l}
\end{equation}
where ${\bf p}_{in}$ is the sum of incident particles momentums,
${\bf p}_{out}$
is that after the collision, $\varepsilon (p)$, $p$ are the energy and
momentum of projectile, ${\bf n}=\frac{{\bf p}}{|p|}$,
${\bf q}$ is momentum transferred ${\bf q}={\bf p}^{\prime }-{\bf p}
$,
$F({\bf q,n})$ is
the scattering amplitude normalized by the condition $\sigma _{tot}=4\pi
ImF(0)$. It
can be obtained from the ordinary amplitude through division by the
wave number of the incident particle.
Amplitude  normalized in such a way is invariant relative to the
Lorentz transforms along the incident particle momentum direction, so, the
$p-d$ amplitude is equal to the $d-p$ amplitude when  the
deuteron moves with the same velocity as
proton at the first case. It is more convenient to speak about $p-d$
amplitude under calculation but we need $d-p$
amplitude for the description of the spin oscillations effect.
Calculating scattering matrix element we use the deuteron wave
function in the following form
\begin{eqnarray}
|_d~{\bf p}>=(2\pi)^{-3/2}
e^{
i{\bf p}\frac{{\bf r}_{1}+{\bf r}_{2}}{2}}
\phi({\bf r}_{1}-{\bf r}_{2})=
\nonumber
\\
(2\pi)^{-3/2}\int\tilde \phi({\bf k})
|_1{\bf k}+\frac{{\bf p}}{2};_2-{\bf k}+\frac{{\bf p}}{2}>d^3
{\bf k}~,
\end{eqnarray}
where $\tilde \phi ({\bf k})=\int \phi ({\bf r})e^{-i{\bf
kr}}d^{3}{\bf r}$, $~~|_{1} {\bf k};_{2}{\bf p}>\equiv |_{1}{\bf
k}>|_{2}{\bf p}>\equiv (2\pi )^{-3/2}e^{i{\bf kr}_{1}}(2\pi
)^{-3/2}e^{i{\bf pr}_{2}}.$ Taking matrix element from (\ref{approx})
we obtain for the single scattering (moving nucleon and rest deuteron are
considered):
\widetext
\begin{eqnarray} <_d{\bf 0}|<{\bf
p}^\prime|\Lambda_1|{\bf p}>|_d~{\bf 0}>=
\frac{1}{(2\pi)^3}\int\tilde\phi^+({\bf k}^\prime) <{\bf
p}^\prime;_2-{\bf k}^\prime;_1{\bf k}^\prime| \Lambda_1|_1{\bf
k};_2-{\bf k};{\bf p}>
\nonumber \\ \times
\tilde\phi({\bf k})d^3{\bf k} d^3{\bf k}^\prime=
\int \tilde\phi^+({\bf k}^\prime)\delta^{(3)}({\bf
k}^\prime-{\bf k}) <{\bf p}^\prime;_1{\bf k}^\prime|\Lambda_1 |_1{\bf
k};{\bf p}>\tilde\phi({\bf k})d^3{\bf k} d^3{\bf k}^\prime
\nonumber \\
=
-\delta^{(3)}({\bf p}^\prime-{\bf p})
\frac{p}{(2\pi)^2\varepsilon(p)}
Sp_{\sigma_1\sigma_2}\{f_1({\bf 0,n})G({\bf 0})\}
\end{eqnarray}
and for the double scattering:
\begin{eqnarray}
<_d{\bf 0}|<{\bf p}^\prime|\Lambda_1\Omega_0\Lambda_2|{\bf p}>
|_d~{\bf 0}>=
\frac{1}{(2\pi)^3}\int\tilde\phi^+({\bf k}^\prime)
<{\bf p}^\prime;_2-{\bf k}^\prime;_1{\bf k}^\prime|
\Lambda_1\Omega_0\Lambda_2|_1{\bf k};
_2-{\bf k}^\prime;{\bf p}>
\nonumber \\
\times
\tilde\phi({\bf k})d^3{\bf k}d^3{\bf k}^\prime=
\frac{1}{(2\pi)^3}\int\tilde\phi^+({\bf k}^\prime)
<{\bf p}^\prime;_1{\bf k}^\prime;|\Lambda_1|_1{\bf k};{\bf q}>
\frac{1}{E-
\varepsilon(q)+i0}
\nonumber \\ \times
<{\bf q};_2-{\bf k}^\prime|\Lambda_2|_2-{\bf
k};{\bf p}>\tilde\phi({\bf k})d^3{\bf q}d^3{\bf k}d^3{\bf k}^\prime
\nonumber
\\ \approx
\frac{\delta^{(3)}({\bf p}-{\bf p}^\prime)}{(2\pi)^7}
\int\tilde\phi^+({\bf k}^\prime)\frac{p}{\varepsilon(p)}f_1({\bf p}-{\bf q}
,{\bf n})
\frac{1}{E-\varepsilon({\bf p}+{\bf k}^\prime-{\bf k})+i0}
\nonumber \\ \times
\frac{p}{\varepsilon(p)}
f_2({\bf q}-{\bf p},{\bf n})\tilde\phi({\bf k})\delta(q_z-k^\prime_z
+k_z-p)
\delta^{(2)}({\bf
q}_\bot-{\bf k}^\prime_\bot+{\bf k}_\bot)d^3{\bf q}
d^3{\bf k}d^3{\bf k}^\prime
\nonumber \\
\approx
\frac{-i\delta^{(3)}({\bf
p}-{\bf p}^\prime)}{(2\pi)^3}\frac{p}{\varepsilon(p)}
Sp_{\sigma_1\sigma_2}\left\{\int d^2{\bf q}_\bot
f_1(-{\bf q}_\bot,{\bf n})f_2({\bf q}_\bot,{\bf n})
]G^{(+)}(2{\bf q})\right\}~,
\label{12}
\end{eqnarray}
\narrowtext
where the form factor $G^{(\pm )}({\bf q}_{\bot })$ is defined as
\begin{eqnarray}
G^{(\pm)}({\bf q}_\bot)=\frac{1}{(2\pi)^4}\int
\tilde\phi({\bf k})
\tilde\phi^+({\bf k}^\prime)
\delta^{(2)}({\bf k}_\bot-{\bf k}_\bot^\prime+\frac{{\bf q_\bot}}{2})
\frac{i}{k_z-k_z^\prime\pm i0}d^3{\bf k}d^3{\bf k}^\prime~,
\\
G({\bf q}_\bot)= G^{(+)}({\bf q}_\bot)+G^{(-)}({\bf q}_\bot),~~~~~~~~~~~~~~~~~~~~~~~~~~~~~
\nonumber
\end{eqnarray}
To derive these equations we have used some assumptions: 1) it is suggested
that $r_{D}>r_{0}$ where $r_{D}$ is the deuteron size and $r_{0}$ is the
interaction size; 2) we remain only the energy of the incident nucleon at the
denominator of $\Omega _{0}$ and bring $\Omega _{0}$ to the eikonal form: $%
(E-\varepsilon ({\bf p}+{\bf k}^{\prime }-{\bf k})+i0)^{-1}=(-\frac{\partial
\varepsilon (p)}{\partial {\bf p}}({\bf k}^{\prime }-{\bf k})+i0)^{-1}=\frac{%
\varepsilon (p)}{p(k_{z}-k_{z}^{\prime }+i0)}~.$ Taking the matrix element
from the remaining terms of (\ref{approx}) and comparing with (\ref{def_l})
we find the forward scattering amplitude of the nucleon on the deuteron:
\widetext
\begin{eqnarray}
F(0)=Sp_{\sigma _{1}\sigma _{2}}\biggl\{(f_{1}(0)+f_{2}(0))G(0)\biggr\}+%
\frac{i}{2\pi }Sp_{\sigma _{1}\sigma _{2}}\biggl\{\int (f_{1}(-{\bf q})f_{2}(%
{\bf q})G^{(+)}(2{\bf q})
\nonumber \\
+f_{2}({\bf q})f_{1}(-{\bf q}))G^{(-)}(2{\bf q}%
)d^{2}{\bf q}\biggl\}  \label{ampl}
\end{eqnarray}
\narrowtext
It is implied in (\ref{ampl}) that ${\bf q}$ is two-dimensional vector.

\section{ Forward elastic scattering amplitude of the deuteron on a nucleon.}

Rewriting (\ref{ampl}) in terms of a profile function $\Gamma ({\bf b})$ and
density $\rho ({\bf r})$ we have for the nucleon deuteron scattering:
\widetext
\begin{eqnarray}
F(0)=\frac{i}{2\pi}\int Sp_{\sigma_1 \sigma_2} \biggl\{ \Gamma_1
({\bf b}-{\bf b}_1)+
\Gamma_2 ({\bf b}-{\bf b}_2)- \Gamma_1 ({\bf b}-
{\bf b}_1) \Gamma_2 ({\bf b}-{\bf b}_2)
\theta (z_1-z_2)- \nonumber \\ \Gamma_2({\bf b}-{\bf b}_2)
\Gamma_1({\bf b}-{\bf b}_1)
\theta (z_2-z_1)\biggr\} \delta ({\bf r}_1-{\bf r}_2) \rho ({\bf r}_1)
d^3{\bf r}_1d^3{\bf r}_2= \nonumber \\
\frac{i}{2\pi}\int Sp_{\sigma_1 \sigma_2} \biggl\{
(\Gamma_1({\bf b}-{\bf b}^{\prime})+
\Gamma_2({\bf b}+{\bf b}^{\prime}))\rho_\bot ({\bf b}^\prime)-
\Gamma_1({\bf b}-{\bf b}^\prime)
\Gamma_2({\bf b}+{\bf b}^\prime)\rho_\bot^{(+)}({\bf b}^\prime) -
\nonumber \\ \Gamma_2({\bf b}+{\bf
b}^\prime)\Gamma_1({\bf b}-{\bf b}^\prime)
\rho_\bot^{(-)} ({\bf b}^\prime)
\biggr\} 
\label{f20} \end{eqnarray}
\narrowtext where $\rho _{\bot }^{(+)}({\bf b})=\int_{0}^{+\infty
}\rho ({\bf r})dz,\quad $ $\rho _{\bot }^{(-)}({\bf b})=
\int_{-\infty }^{0}\rho ({\bf r})dz,\quad $ $\rho _{\bot }({\bf b}%
)=\rho _{\bot }^{(+)}({\bf b})+\rho _{\bot }^{(-)}({\bf b%
}),\quad $ ${\bf r}\equiv \{{\bf b},z\}$. $\rho ({\bf r)}$ is   the
nucleon density in deuteron and spin density matrix of deuteron nucleons
simultaneously. The expression (\ref{f20}) implies that, if the incident
particle firstly scatters on the nucleon 1 it spin wave function
is acted by $ \Gamma _{1}$ , and then by the $\Gamma _{2}$. If the
first collision happens with nucleon 2 the $\Gamma _{2}$ acts
firstly. Profile-function $\Gamma ( {\bf b)}$ connected with the
$N-N$ nucleon scattering amplitude by the relation: $f({\bf
q})=\frac{i}{2\pi }\int \Gamma ({\bf b})e^{-i{\bf qb}}d^{2} {\bf b}$
and form factors connected with the density as $G^{(\pm )}({\bf q}
)=\int \rho _{\bot }^{(\pm )}({\bf b})e^{i{\bf q}
{\bf b}}d^{2}{\bf b}$. If we look at the expression (\ref
{ampl}) we see that in the resulting expression the integration area over
transferred momentum  $q$ is restricted by the deuteron form factor. In
supposition of a sharper dependence of deuteron form factor on ${\bf q}$ we
may carry out N-N scattering amplitude from the integration or
using the model profile function:
\begin{eqnarray}
\Gamma _{\alpha }({\bf b}) &=&\frac{2\pi }{i}\biggl\{a_{\alpha }+v_{\alpha }(
\vec{\sigma}\vec{\sigma}_{\alpha })+e_{\alpha }(\vec{\sigma}_{\alpha }{\bf n}
)(\vec{\sigma}{\bf n})-\frac{c_{\alpha }}{m}(\vec{\sigma}  \nonumber \\
&&+\vec{\sigma}_{\alpha })\times {\bf n}\frac{\partial }{\partial {\bf b}}-
\frac{d_{\alpha }}{m^{2}}(\vec{\sigma}_{\alpha }\frac{\partial }{\partial
{\bf b}})(\vec{\sigma}\frac{\partial }{\partial {\bf
b}})\biggr\}\delta^{(2)} ( {\bf b}).  \label{model} \end{eqnarray}
wher $\delta^{(2)} ({\bf b})$ is the two dimensional Dirac delta-function,
$m$ is the
nucleon mass.  This profile function corresponds to the scattering
amplitude \begin{eqnarray} f_{\alpha }({\bf q}) &=&a_{\alpha
}+v_{\alpha } (\vec{\sigma}_{\alpha }\vec{\sigma})+e_{\alpha }(%
\vec{\sigma}_{\alpha }{\bf n})(\vec{\sigma}{\bf n})+  \nonumber \\
&&\frac{ic_{\alpha }}{m}(\vec{\sigma}_{\alpha }+\vec{\sigma}){\bf q}\times
{\bf n}+\frac{d_{\alpha }}{m^{2}}(\vec{\sigma}_{\alpha }{\bf q})(\vec{\sigma}{\bf q})~,
\end{eqnarray}
where $a_{\alpha },~v_{\alpha }~...$ do not depend on ${\bf q}$.
Considering T and P invariance  we can represent $\rho ({\bf r)}$
as: \widetext
\begin{eqnarray}
\rho ({\bf r}) =\frac{1}{4}\{A_{0}+A_{1}{\bf S}\left( \vec{\sigma}_{1}+\vec{%
\sigma}_{2}\right) +A_{2}({\bf S}{\bf r})(\vec{\sigma}_{1}+\vec{\sigma}_{2})\cdot
r+A_{3}({\bf S}{\bf r})^{2}+A_{4}(\vec{\sigma}_{1}\vec{\sigma}_{2})
\nonumber \\
+A_{5}(\vec{
\sigma}_{1}{\bf r})(\vec{\sigma}_{2}{\bf r})+A_{6}((\vec{\sigma}_{1}{\bf S})(\vec{\sigma}%
_{2}{\bf S})
+(\vec{\sigma}_{2}{\bf S})(\vec{\sigma}_{1}{\bf S}))+A_{7}(\vec{\sigma}%
_{1}{\bf r})(\vec{\sigma}_{2}{\bf r})({\bf S}{\bf r})^{2}
\nonumber \\
+A_{8}(\vec{\sigma}_{1}\vec{\sigma}%
_{2})({\bf S}{\bf r})^{2}+A_{9}((\vec{\sigma}_{1}\times {\bf S}\cdot{\bf  r})(\vec{\sigma%
}_{2}\times {\bf S}\cdot{\bf  r})
+(\vec{\sigma}_{2}\times {\bf S}\cdot{\bf  r})(\vec{\sigma}_{1}\times {\bf S}%
\cdot {\bf r}))
\nonumber \\
+A_{10}((\vec{\sigma}_{1}{\bf r})(\vec{\sigma}_{2}{\bf S})({\bf S}{\bf r})+(%
\vec{\sigma}_{1}{\bf S})(\vec{\sigma}_{2}{\bf r})({\bf S}{\bf r})+(\vec{\sigma}_{1}{\bf r})(%
{\bf S}{\bf r})(\vec{\sigma}_{2}{\bf S})+({\bf S}{\bf r})(\vec{\sigma}_{2}{\bf r})(\vec{\sigma}%
_{1}{\bf S}))\}  \label{rho}
\end{eqnarray}
\narrowtext
 $A_{n}$ are real functions of $r$ everywhere in (\ref{rho}). $A_{0}-A_{10}$
are found from the deuteron wave function:
\begin{equation}
\phi _{m}=\frac{1}{\sqrt{4\pi }}\left( \frac{U(r)}{r}+\frac{1}{\sqrt{8}}%
\frac{W(r)}{r}S_{12}\right) \chi _{m},
\end{equation}
where $U(r)$ is the radial deuteron S-wave function and $W(r)$ is the radial
D-function, $\chi _{m}$ is the spin wave function of  two nucleons
with spin projection $m$ , $S_{12}=\frac{3(\vec{\sigma}_{1}{\bf
r})(\vec{\sigma}_{2}{\bf r})}{r^2}-(\vec{\sigma} _{1}\vec{\sigma}_{2})$.
Taking into account $\rho ({\bf r})=8\phi (2{\bf r})\phi ^{+}(2{\bf r}) $ we find nucleon
density matrix for the deuteron  at the state corresponding
to spin projection 1:
\widetext \begin{eqnarray} \langle 1 &\mid &\rho ({\bf r}))\mid 1\rangle
=\frac{1}{\pi }\left( \frac{U(2r)}{r}
+\frac{1}{\sqrt{8}}\frac{W(2r)}{r}S_{12}\right) \chi _{1}\chi _{1}^{+}\left(
\frac{U(2r)}{r}+\frac{1}{\sqrt{8}}\frac{W(2r)}{r}S_{12}\right)  \nonumber \\
&=&\frac{1}{\pi }\left( \frac{U(2r)}{r}+\frac{1}{\sqrt{8}}\frac{W(2r)}{r}%
S_{12}\right) \frac{(1+\vec{\sigma}_{1}\cdot {\bf e})}{2}\frac{(1+\vec{\sigma%
}_{2}\cdot {\bf e})}{2}\left( \frac{U(2r)}{r}+\frac{1}{\sqrt{8}}\frac{W(2r)}{%
r}S_{12}\right) ,
\end{eqnarray}
\narrowtext
where ${\bf e}$ is the unit vector in the deuteron spin direction. From the
other side $\langle 1\mid \rho \mid 1\rangle $ can be obtained by taking
the matrix element from $(\ref{rho})$ for the deuteron state corresponding
to spin projection 1. It can be obtained  by comparing the expressions :

\noindent $A_0(r)=u^2-8uw+16w^2,\;\;A_1=u^2-2uw-8w^2,
\;\;
r^2A_2=6uw+12w^2,\;\;r^2A_3=12uw-12w^2,$

\noindent $A_4=8w^2+8uw-u^2,\;\;r^2A_5=-24w^2,
\;\;
A_6=u^2-2uw+4w^2,\;\;r^4A_7=72w^2,$

\noindent $r^2A_8=-12w^2,\;\;r^2A_9=-6uw,\;\;r^{2}A_{10}=-12w^{2}.\;\;$

Functions $u$ and $w$ are expressed through $U$ and $W$ by: $W(2r)=4\sqrt{\pi }%
rw(r) $, $U(2r)=\sqrt{2\pi }ru(r)$. By substituting the
expression for $\rho (%
{\bf r})$ and for the profile-function to the (\ref{f20}) it is possible to
calculate the forward $p-d$ scattering amplitude (Appendix) that
is also $d-p$ scattering amplitude for the two times
larger energy due to normalization used. The values $a,b,c...$ have been
taken from the SAID
\cite{arnd}
phase shift analysis
(solution SP98)
and Nijmegen deuteron S- D- wave function \cite{stoks} results have
been obtained which are pictured on the figure \ref{low}. We see that the
contribution of the spin dependent N-N interactions is small for the
deuteron energy greater than $5~Gev$. Note, that the approximation of
the constant $a$,$v$,$e$,$c$,$d$ used in (\ref{model}) is very rough,
because calculations show that spin-dependent deuteron form factors
have less sharper $q$ dependence than spin-less form factor.
Consider, for instance, form factor with neglecting the dependence on spin
of deuteron nucleons.
\begin{eqnarray} G({\bf q}) &=&G_{0}({\bf
q})+({\bf Sn})^{2}G_{s}({\bf q}) \\ &=&\int
(A_{0}(r)+A_{3}(r)b^{2})e^{i{\bf qb}}d^{2}{\bf b}dz+
({\bf Sn})^{2}\int A_{3}(r)(z^{2}-\frac{b^{2}}{2})e^{i{\bf qb}}d^{2}{\bf b}%
dz.  \nonumber
\end{eqnarray}
Form factors $G_{0}({\bf q})$ and $G_{s}({\bf q})$ are shown in
fig. \ref{form} and
we see that the $G_{s}({\bf q})$ is wider than ordinary $G_{0}(%
{\bf q})$ considered . We still use above the approximation  to simplify
calculations.

\section{Deuteron proton scattering amplitude at high energies.}

It has been found above that
we may neglect the spin dependence of the N-N scattering amplitude
at high energies . In this case the
$T_2$ term of the
proton deuteron forward scattering amplitude is
written down as:
\begin{equation}
T_2=\frac{i}{2\pi }\int f(t)f(t)G_s(2{\bf q})d^{2}{\bf q}~,
\label{51}
\end{equation}
where $t=-{\bf q}^{2}$.
 We take N-N scattering amplitude from the \cite{kob1,kob2} (dipole
pomeron is considered there):
\begin{equation}
f(s,t)=P(s,t)+\Phi (s,t)\pm \omega (s,t),  \label{rfp}
\end{equation}
where $"+"$ sign corresponds to the $p\bar{p}$ scattering and $"-"$ to the $%
pp$ one,  $s$ is the nucleons energy squared in their
center-of-mass frame. The amplitude contains dipole pomeron contribution
\begin{equation}
P(s,t)=ig^{2}\ln (s\;e^{-\frac{i\pi }{2}}/s_{1})(s\;e^{-\frac{i\pi }{2}%
}/s_{2})^{\alpha _{P}(t)-1}\exp (-b_{P}t),
\end{equation}
$\Phi $ reggeon
\begin{equation}
\Phi (s,t)=ir_{\Phi }(s\;e^{-\frac{i\pi }{2}}/s_{0})^{\alpha _{\Phi
}(t)-1}\exp (-b_{\Phi }t)
\end{equation}
and $\omega $ reggeon
\begin{equation}
\omega (s,t)=r_{\omega }(1-t/t_{\omega })(s\;e^{-\frac{i\pi }{2}%
}/s_{0})^{\alpha _{\omega }(t)-1}\exp (-b_{\omega }t)
\end{equation}
contributions. Parameters used are listed in \cite{kob1,kob2}. The N-N
scattering amplitude (\ref{rfp}) obeys the derivative dispersion
relation (DDR)\cite{bron,jac,block}.

Let us remember how the DDR arises.
The requirement of analyticity is that the $pp$ and $\bar pp$ elastic
scattering are described by the single analytic
amplitude function ${\mathfrak f}(s,t)$ \cite{eden}:
\begin{equation}
{\mathfrak f}_{p\bar{p}}(u+i0,t)={\mathfrak f}_{pp}(4m^{2}-u-i0,t)~,
\end{equation}
where the relation $s+u+t=4m^{2}$ is used. For the large $u$ we have:
\begin{equation}
{\mathfrak f}_{p\bar{p}}(u+i0,t)={\mathfrak f}_{pp}(-u-i0,t)=
{\mathfrak f}_{pp}^{\ast }(-u+i0,t)~.
\end{equation}
In the latest equality  ${\mathfrak f}_{pp}(u)$ is assumed to be a
real function of $u$.  Changing $u$ by $s$ we have \begin{equation}
{\mathfrak f}_{p\bar{p}}(s+i0,t)={\mathfrak f}_{pp}^{\ast }(-s+i0,t)
\end{equation}
Constructing combinations ${\mathfrak f}_{+}=
\frac{{\mathfrak f}_{p\bar{p}}+
{\mathfrak f}_{pp}}{2}$ and ${\mathfrak f}_{-}=\frac{{\mathfrak f}_{pp}-{%
\mathfrak f}_{p\bar{p}}}{2}$ we find
\begin{equation}
{\mathfrak f}_{+}(se^{i\pi })={\mathfrak f}_{+}^{\ast }(s)~,~~~~~~~~~{%
\mathfrak f}_{-}(se^{i\pi })=-{\mathfrak f}_{-}^{\ast }(s)~,
\end{equation}
and for  our amplitude normalization $f\sim \frac{{\mathfrak f}}{s}$ :
\begin{equation}
f_{+}(se^{i\pi })=-f_{+}^{\ast }(s)~~~~~~~~~~f_{-}^{\ast }(se^{i\pi
})=f_{-}^{\ast }(s)~.
\label{510}
\end{equation}
Solution of this relations are $f_{+}(s)=iK(se^{-\frac{i\pi }{2}})$ and $%
f_{-}(s)=K(se^{-\frac{i\pi }{2}})$, where $K(s)$ is a real function $K^{\ast
}(s)=K(s^{\ast })$. It is easy to check that $f_{+}(s)$ and $f_{-}(s)$ are
satisfied to the dispersion relations:
\begin{eqnarray}
Re\;f_{+}(s,t) &=&tg\left( \frac{\pi }{2}\frac{d}{d\ln (s)}\right)
Im\;f_{+}(s,t)~~  \nonumber \\
Im\;f_{-}(s,t) &=&-tg\left( \frac{\pi }{2}\frac{d}{d\ln (s)}\right)
Re\;f_{-}(s,t)~.
\end{eqnarray}
The difference between the particle and antiparticle cross sections goes to zero at
high energies \cite{kob2} and the single DDR for $f$  remains. The $T_2$ term of
the forward deuteron -proton scattering amplitude
satisfies the DDR at high energies
\begin{equation}
Re\;T_{2}(s)=tg\left( \frac{\pi }{2}\frac{d}{d\ln (s)}\right)
Im\;T_{2}(s)~~
\end{equation}
because if $f$ satisfies the first equation of (\ref{510}),
$T_2$ in (\ref{51}) has the same form as $f$ ($G(q)$ is a real quantity).
Certainly we can deduce
this by straight way by considering $dp$ and $d\bar{p}$ channels. So
measuring real part through the effect of the deuteron spin oscillation
we have a new test for  DDR checking. Oscillation phase
(rotational angle) and dichroism calculated with the amplitude
(\ref{rfp}) is shown in the Fig. \ref{high}.  We see that  rotation angle
goes through zero in vicinity of $E \sim 1~Tev$ and then its absolute
value increases with energy.

\section{Cancelation of the Coulomb interaction influence on
spin oscillation and rotation phenomena.}

It is necessary to discuss the influence of electromagnetic interaction.
Charged particle
beam moving in a matter is under the action of Coulomb multiple
scattering.
Since  the Coulomb effects at high energies are
significant only at small angles, it is enough  to
consider  the terms proportional to $q^{-2}$ and to $q^{-1}$.
Divergent at small $q$ terms
of the deuteron electromagnetic scattering
amplitude on an unpolarized nuclei of charge $Z$ can be written as
(another terms which is convergent at small $q$ is
much  smaller than the nuclear part of the scattering amplitude):
\begin{equation}
F_{col}=a_{col}(q)-\frac{i c_{col}(q)}{m_d}(
{\bf S}\cdot{\bf n}\times{\bf q})~,
\label{col}
\end{equation}
where
$a_{col}(q)=-\frac{2Z\alpha}{q^2}\frac{\gamma_d}{\sqrt{\gamma^2_d-1}}~~,$
$c_{col}=\frac{Z\alpha}{q^2}\left(\mu_d -
\frac{\gamma_d}
{
\gamma_d+\frac{m_d}{m}+
\sqrt{1+\left(\frac{m_d}{m}\right)^2+
2\frac{m_d}{m}\gamma_d}
}  \right)~,
$ $\alpha=1/137$ is the fine structure constant, $\gamma_d$ is
the deuteron Lorentz factor,
$m_d$ and $m$ are the deuteron and target nuclei masses
correspondingly.
The nuclear
part of  scattering amplitude has smooth $q$--behavior
and much smaller
than the Coulomb one at small $q$.
Coulomb scattering results in the particle
beam spreading over small angle but
still it does not prevent spin rotation and oscillation if to
speak about rotation of the polarization averaged
over beam spreading angle.
This problem is discussed earlier in terms of
density matrix formalism
\cite{cher,shek}
and it has been shown that
the second term of the electromagnetic amplitude results in only
insignificant depolarization of the beam
lowering absolute value of the beam polarization $|\xi|$
\begin{equation}
\frac{d\xi^2}{d z}=-4\pi\rho z
\frac{\overline{\mid c_{col}\mid^2}}{2m_d^2}\xi^2~,
\end{equation}
where
\begin{eqnarray}
\overline{
|c_{col}|^2}=
\int\limits^{q_{max}}_{q_{min}}\mid c_{col}\mid^2 q^3dq
=
(Z\alpha)^2
~~~~~~~~~~~~~~~~~
\nonumber
\\
\times
\left(\mu_d -
\frac{\gamma_d}
{
\gamma_d+\frac{m_d}{m}+
\sqrt{1+\left(\frac{m_d}{m}\right)^2+
2\frac{m_d}{m}\gamma_d}
}  \right)
\ln\left(\frac{q_{max}}{q_{min}}\right)~.
\nonumber
\end{eqnarray}

Maximum momentum transferred
$q_{max}=\frac{2\pi}{r_d}$
is restricted by the deuteron
size $r_d$ or by the collimation angle $\theta_{det}$ of the detector
$q_{max}=\theta_{det} k_d$, where $k_d$ is the deuteron momentum.
Minimal momentum transferred
$q_{min}=\frac{2\pi}{r_b}$
is determined by the
screening radius approximately
equaled to the radius of the first Born orbit $r_b$.
First term of the amplitude (\ref{col}) results in the correction due to
Coulomb nuclear interference which can be taken into account
\cite{shek}
by
changing $F(0)\rightarrow F(0)\exp(i\bar a_{col})$ in
(\ref{ind_ref1})
, where
\[
\bar a_{col}=\int\limits^{q_{max}}_{q_{min}}
a_{col}(q)qdq=-2 Z\alpha\frac{\gamma_d}{\sqrt{\gamma_d^2-1}}~.
\]
However,
nuclear amplitude also contains an electromagnetic correction due to
Coulomb
wave distortion which turns out to be
$F(0)=F_{nucl}\exp(-i\bar a_{col}))$
(Bethe's phase
\cite{beth,sol,rix,losh,islam,west,savr,kaf,gorsh,sol2} ),
where $F_{nucl}$ is "pure" nuclear scattering amplitude
under the turned off electromagnetic interaction.
We see that these corrections cancel each other and "pure"
nuclear amplitude can be measured with the help of spin
oscillation and rotation and dihroism phenomenon.

\section{Deuteron scattering by nuclei.}

 Deuteron scattering by nuclei should be considered on
the basis of the Glauber multiscattering theory with Gribov
inelastic corretions \cite{gribov,karm}.
However, we give simplified analysis of the discussed effect.
Notice, that two types of the diagrams of deuteron double
sattering exist \cite{cz}.
For the first type diagrams both deuteron
nucleons strike the same nucleon of the nucleus.
For the second type diagrams deuteron nucleons are scattered by
different nucleons of a nucleus (figure \ref{diag}).

  Contribution of the first type diagrams
with screening can be expressed through
the deuteron-nucleon scattering amplitude. Let us write
deuteron-nucleus scattering amplitude as
\begin{equation}
F_{dA}=\frac{i}{2\pi}\int
\left(1-e^{-\Omega_{dA}({\bf b})}\right)d^2{\bf b}~~,
\label{dA}
\end{equation}
where eikonal function is expressed through the deuteron-nucleon
forward scattering amplitude $F$, the nucleus radius $R$, the deuteron
radius $r_d$ and the atomic mass number $A$:
\begin{equation}
\Omega_{dA}=\frac{3iFA}{R^2+r_d^2}
\exp\left(-\frac{3b^2}{2(R^2+r_d^2)} \right)~~.
\label{omdA}
\end{equation}
Substituting (\ref{omdA}) to the (\ref{dA}) we find
\begin{equation}
F_{dA}=i\frac{R^2+r_d^2}{3}
Ein\left(-\frac{3iFA}{R^2+r_d^2}  \right)~,
\end{equation}
where exponential integral \cite{Abram} function $Ein$ is
$Ein(z)=\int\limits^{z}_{0}(1-e^{-t})dt/t~,~~~Re~ z>0$.
Remembering that the values of deuteron-nucleon scattering amplitudes
$F$ for the deuteron spin projectionson the momentum
$S_z=0$ and $\mid S_z\mid=1/2$ are
slightly  different  we express the $T_2$
term of the forward deuteron-nucleus scattering amplitude through
the $T_2$-term of the deuteron-nucleon amplitude
\begin{equation}
T_2^{dA}=\frac{\partial F_{dA}}{\partial F} T_2
=i\frac{R^2+r_d^2}{3}
\left(1-exp\left(\frac{3iFA}{R^2+r_d^2}\right)\right)\frac{T_2}{F}
\label{otvet}
\end{equation}
Taking   $R=1.12\times A^{1/3}~~fm $
we conclude  that $T_2^{dA}$ is proportional to $A$ at small
$FA/(R^2+r_d^2)$ and
to the $A^{2/3}$ at large one.

Second type diagrams can be described through the elastic scattering
amplitude of the nucleon by nucleus.
Nucleon-nucleus eikonal has the same form as (\ref{omdA})
$\Omega_{NA}=\frac{3ifA}{R^2}\exp\left(-\frac{3b^2}{2R^2}\right)$ and
connected with the N-A amplitude $f_{NA}(q)$ as in (\ref{dA}).
$T_2$ term of
d-A forward scattering amplitude is written down as
\begin{equation}
T_2^{dA}=\frac{i}{\pi}\int f_{NA}(q)f_{NA}(-q)G_s(2q)d^2{\bf q}.
\label{ee}
\end{equation}
We can approximate $G_s(q)
\approx {\mathfrak a}_s q^2 $ and substituting to
(\ref{ee}) we find
\begin{equation} T_2^{dA}=-\frac{i {\mathfrak
a}_s}{4}\left( \exp\left( \frac{6ifA}{R^2}\right)
+Ein\left(-\frac{6ifA}{R^2} \right)  \right)
\end{equation}
Assimptotic
$$
Ein(z)\approx \cases{z-\frac{z^2}{2}, & $\mid z \mid \ll 1$; \cr
-0.5772
+ \ln(z),& $ \mid z \mid \gg 1$ }
$$
shows that  $T_2^{dA}\sim A^{2/3}$ at small
$fA/R^2$  and
$T_2^{dA}\sim ln(A)$ at large one.
Relative contribution of the diagrams of the first and the second type
is seen from in Fig. \ref{adep}.

\section{Conclusion}

Let us estimate the effect value for  hydrogen
target of $0.0675~g/cm^{3}$ density. The   oscillation phase
and dihroism ${\mathcal A}$ are described by expressions:
\begin{equation}
\phi (z)=2\pi \rho Re(F_{1}(0)-F_{0}(0))z,
\label{fase}
\end{equation}
and
\begin{equation}
{\mathcal A}(z)={\frac{I_{0}(z)-I_{\pm 1}(z)}{I_{0}(z)+I_{\pm 1}(z)}}=2\pi \rho
Im(F_{1}(0)-F_{0}(0))z~,
\label{dih}
\end{equation}
where $I_{m}(z)$ is the intensity
 of deuterons  with the spin projection   $m$ on the depth $z$
if the incident beam is unpolarized.
For an  unpolarized
beam    $I_{-1}(0)=I_{0}(0)=I_{1}(1)=I/3$.
It is obtained for hydrogen target
$\phi \sim 1.1\times 10^{-2}~~{{rad}/{m}},~~~$
${\mathcal A}\sim -3.9\times 10^{-2}~~m^{-1}~~$
at the deuteron laboratory energy $1~Gev$  (Fig. \ref{low})
and
$\phi \sim -5.2\times 10^{-4}~~{{rad}/{m}},~~~$
${\mathcal A}\sim -2.3\times 10^{-3}~~m^{-1}~$
at $E_{lab}=10~Tev$ (Fig. \ref{high}).

For carbon target of  $2.2~g/cm^3$ density
  we find with the help of  (\ref{otvet})
$\phi \sim 0.22~~{{rad}/{m}},~~~$
${\mathcal A}\sim -0.65~~m^{-1}~~$
at the deuteron laboratory energy $1~Gev$
and
$\phi \sim -5.8\times 10^{-4}~~{{rad}/{m}},~~~$
${\mathcal A}\sim -3.1\times 10^{-2}~~m^{-1}~$
at $E_{lab}=10~Tev$.

Note, that  above  $1~ Tev$ the absolute value of the spin rotational
anlge grows asymptotically with the energy (Fig. \ref{abs}).  Thus
the effect of the deuteron spin oscillations and dichroism allow
measurement of the $T_{2}$ term of the deuteron elastic scattering amplitude
 over a broad energy range. This gives information about
rescattering of  nucleons, deuteron wave function, spin-dependent
scattering amplitude of  nucleons, high energy dependence of the
real part of N-N scattering amplitude.  \widetext

\newpage
\section{Appendix}

The expression for the forward scattering amplitude of the deuteron on
a proton has the form
\begin{eqnarray}
F(0) &=&a_{1}+a_{2}+\{(v_{1}+v_{2})(\vec{\sigma}{\bf S})+(e_{1}+e_{2})(%
\vec{\sigma}{\bf n})({\bf S}{\bf n})\}4\pi ^{2}\int\limits_{0}^{+\infty
}A_{1}(z)z^{2}dz  \nonumber   \\
&&+\pi i\int\limits_{0}^{+\infty }\biggl\{%
\{a_{1}a_{2}A_{0}+v_{1}v_{2}(3A_{4}+z^{2}A_{5}+4A_{6}+4z^{2}A_{9})+(e_{1}e_{2}+v_{1}e_{2}
\nonumber \\
&&+v_{2}e_{1})(A_{4}+z^{2}A_{5})-\frac{1}{m^{2}}(d_{2}v_{1}+v_{2}d_{1})(%
\frac{A_{4}^{\prime }}{2z}+\frac{3}{2}A_{5}+\frac{A_{6}^{\prime }}{z}%
+zA_{9}^{\prime }  \nonumber \\
&&+A_{8}+6A_{10})-\frac{c_{1}c_{2}}{2m^{2}}(\frac{A_{0}^{\prime }}{z}+\frac{%
A_{4}^{\prime }}{z}-A_{5}+\frac{2A_{6}^{\prime }}{z}+2A_{8}+2zA_{9}^{\prime
}-4A_{10}  \nonumber \\
&&+4A_{3})\}+({\bf S}{\bf n})^{2}%
\{a_{1}a_{2}z^{2}A_{3}+v_{1}v_{2}(z^{4}A_{7}+3z^{2}A_{8}-2z^{2}A_{9}
\nonumber \\
&&+4z^{2}A_{10})+(e_{1}e_{2}+v_{1}e_{2}+v_{2}e_{1})(2A_{6}+z^{4}A_{7}+z^{2}A_{8}+4z^{2}A_{10})-%
\frac{1}{4m^{2}}(d_{2}v_{1}  \nonumber \\
&&+v_{2}d_{1})(-\frac{2A_{6}^{\prime }}{z}+6z^{2}A_{7}+2zA_{8}^{\prime
}-2A_{8}-2zA_{9}^{\prime }-4A_{9}-12A_{10})  \nonumber \\
&&+\frac{c_{1}c_{2}}{2m^{2}}(-zA_{3}^{\prime }+2A_{3}+\frac{A_{6}^{\prime }}{%
z}+z^{2}A_{7}-zA_{8}^{\prime }+A_{8}+zA_{9}^{\prime }-6A_{9}-2A_{10})\}
\nonumber \\
&&+(\vec{\sigma}{\bf S})\{(a_{1}v_{2}+v_{1}a_{2})A_{1}-\frac{1}{4m^{2}}%
(a_{1}d_{2}+a_{2}d_{1})(\frac{A_{1}^{\prime }}{z}+3A_{2})  \nonumber \\
&&-\frac{1}{2m^{2}}c_{1}c_{2}(\frac{A_{1}^{\prime }}{z}-A_{2})-\frac{i}{2m}%
(c_{1}v_{2}+v_{1}c_{2}+c_{1}e_{2}+e_{1}c_{2})zA_{2}\}  \nonumber \\
&&+(\vec{\sigma}{\bf n})({\bf Sn}%
)\{(a_{1}v_{2}+v_{1}a_{2})z^{2}A_{2}+(a_{1}e_{2}+e_{1}a_{2})(A_{1}+z^{2}A_{2})
\nonumber \\
&&+\frac{1}{4m^{2}}(a_{1}d_{2}+a_{2}d_{1})(\frac{A_{1}^{\prime }}{z}+3A_{2})+%
\frac{1}{2m^{2}}c_{1}c_{2}(\frac{A_{1}^{\prime }}{z}-A_{2})  \nonumber \\
&&+\frac{i}{2m}(3c_{1}v_{2}+3v_{1}c_{2}+c_{1}e_{2}+e_{1}c_{2})zA_{2}\}%
\biggr\}dz~,
\end{eqnarray}
\narrowtext
where $a,v,c,e,d$ are constant and $A_{n}$ are function of $z$, prime means
differentiation on $z$.

\bigskip
\begin{figure}[tbp]
\epsfxsize=9.0 cm
\epsfbox[0 0 230 320]{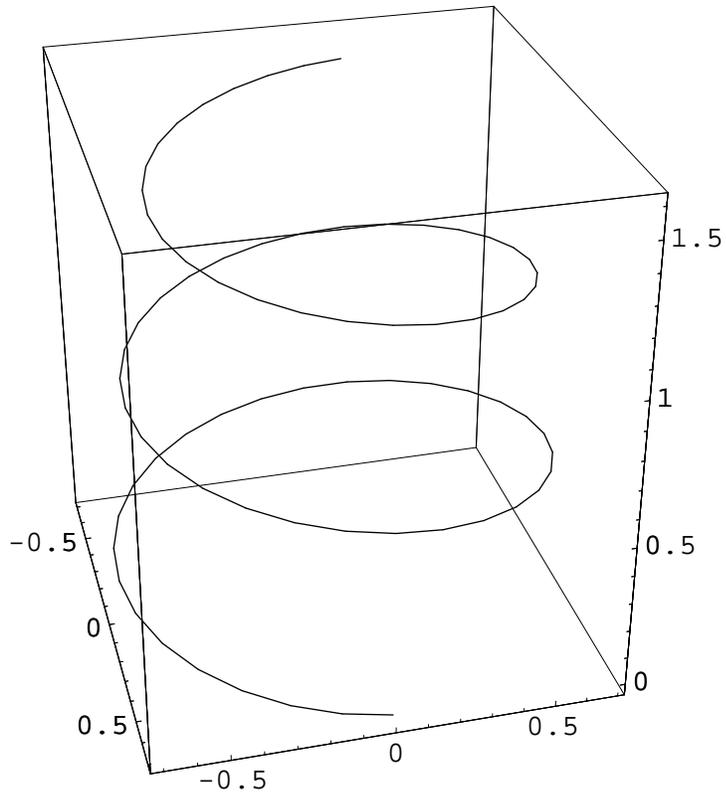}
\caption{The example of spin rotation of the particle with spin S=1 moving
in unpolarized medium.}
\label{rotation}
\end{figure}

\begin{figure}[htbp]
\epsfxsize=9.0 cm
\epsfbox[0 280 330 620]{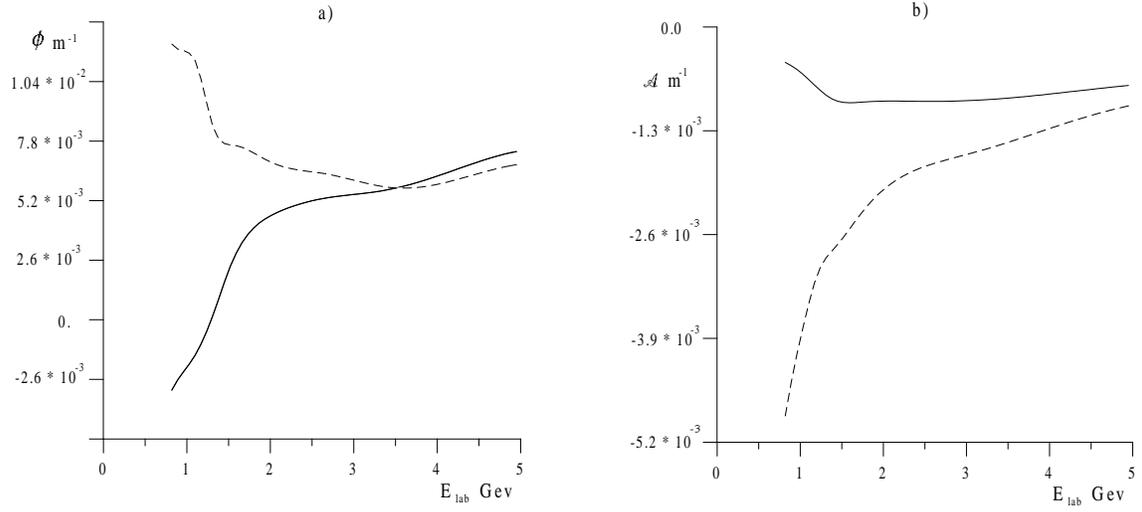}
\caption{Spin rotation angle and dichroism in hydrogen target
before 5 Gev.
Solid curve correspounds to the calculation with
the spinless $N-N$ amplitudes.}
\label{low}
\end{figure}
\begin{figure}[htbp]
\epsfxsize=7.0 cm
\epsfbox[0 150 330 920]{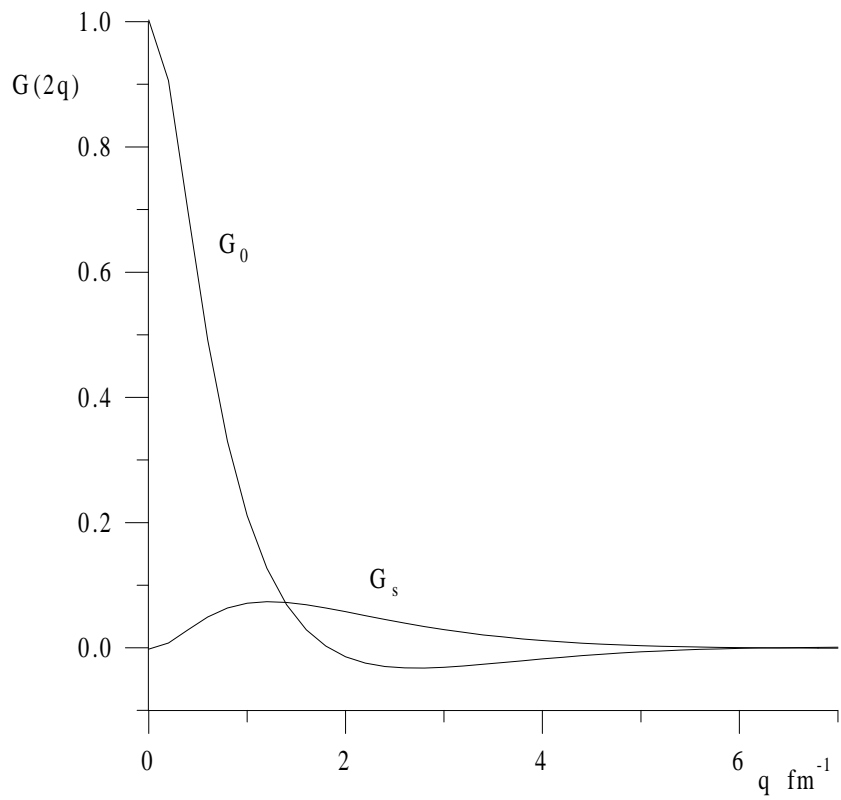}
\caption{Deuteron spinless and quadroupole form factors.}
\label{form}
\end{figure}

\begin{figure}[htbp]
\epsfxsize=9.0 cm
\epsfbox[0 280 330 620]{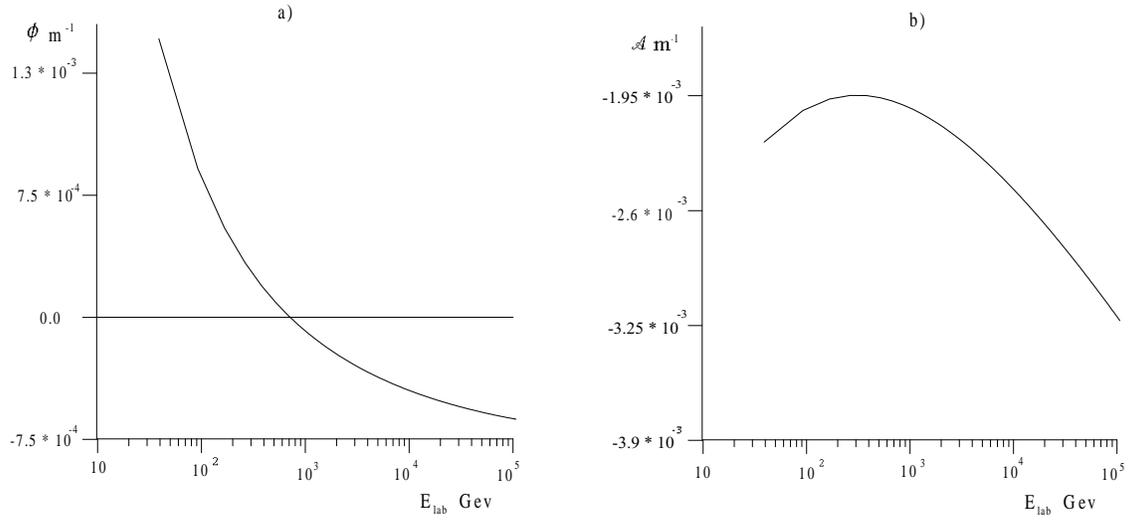}
\caption{
Rotation angle, dichroism in hydrogen target at
high energies.
 }
\label{high}
\end{figure}
\begin{figure}[htbp]
\vspace{-1. cm}
\setlength{\unitlength}{0.15in}
\begin{picture}(33,33)(0,0)
\put(6.,15){${\bf.~~~.~~~.}$}
\put(2,14){\line(1,0){11}}
\put(2,16){\line(1,0){11}}
\put(2,17  ){\line(1,0   ){11  }}
\put(2,18  ){\line(1,0   ){3  }}
\put(2,24   ){\line(1,0   ){3  }}
\put(2,25   ){\line(1,0   ){5.67  }}
\put(5,18  ){\line(1,3   ){2  }}
\put(5,24   ){\line(1,-3   ){1  }}
\put(6,21   ){\line(1,0   ){3  }}
\put(9,21  ){\line(1,3   ){1.33  }}
\put(7.67,25   ){\line(1,-3   ){2.33  }}
\put( 10,18  ){\line(1,0   ){3 }}
\put( 7,24  ){\line(1,0   ){ 6 }}
\put(10.33,25  ){\line(1,0   ){2.67  }}
\put(1,24.5 ){$d$}
\put(1,14){$A$}
\put(1,16 ){$3$}
\put(1,17 ){$2$}
\put(1,18 ){$1$}
\put(7.5,21.5){$1$}
\put(7.5,27){I}
\put(6,21){\circle*{0.4}}
\put(9,21){\circle*{0.4}}
\put(21.,15){${\bf.~~~.~~~.}$}
\put(17,14){\line(1,0){11}}
\put(17,16){\line(1,0){11}}
\put(17,17){\line(1,0   ){5.67  }}
\put(17,18  ){\line(1,0   ){3  }}
\put(17,24   ){\line(1,0   ){3  }}
\put(17,25  ){\line(1,0   ){5.67  }}
\put(25.33,25){\line(1,0   ){2.67}}
\put(20,18  ){\line(1,3   ){2  }}
\put(20,24  ){\line(1,-3   ){2  }}
\put(22.67,17   ){\line(1,3   ){2.67  }}
\put(22.67,25  ){\line(1,-3   ){2.67  }}
\put( 22,18 ){\line(1,0   ){6  }}
\put( 22,24 ){\line(1,0   ){ 6 }}
\put(25.33,17   ){\line(1,0   ){2.67  }}
\put(16,24.5 ){$d$}
\put(16,16 ){$3$}
\put(16,17 ){$2$}
\put(16,18 ){$1$}
\put(16,14){$A$}
\put(21,21){\circle*{0.4}}
\put(24,21){\circle*{0.4}}
\put(22.5,27){$\amalg$}
\end{picture}
\vspace{-5. cm}
\caption{
Diagrams corresponding to the two kinds of the deuteron nucleons
double scattering by nucleus.
}
\label{diag}
\end{figure}
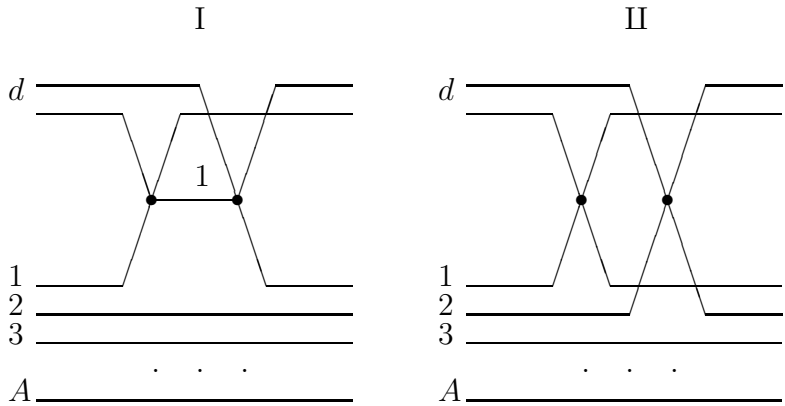
\begin{figure}[htbp]
\epsfxsize=7.0 cm
\epsfbox[0 150 330 920]{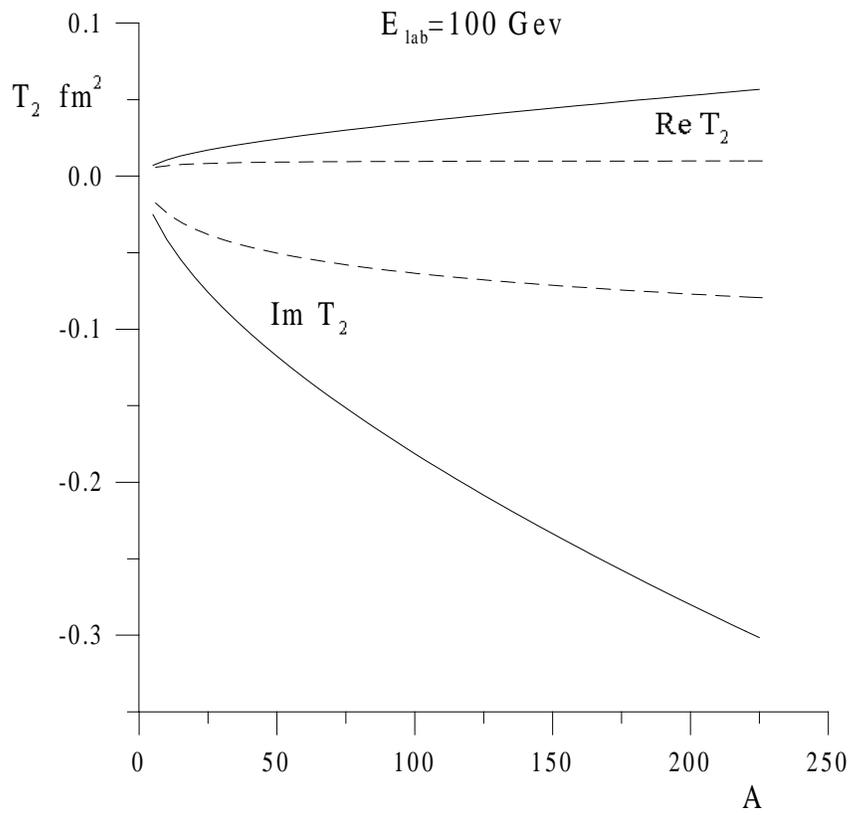}
\caption{
Dependence of the $T_2$-term   from atomic wedth
for the fist  (solid line) and for the
second (dashed line) kinds of diagrams.
}
\label{adep}
\end{figure}

\begin{figure}[tbp]
\epsfxsize=7.0 cm
\epsfbox[0 150 330 920]{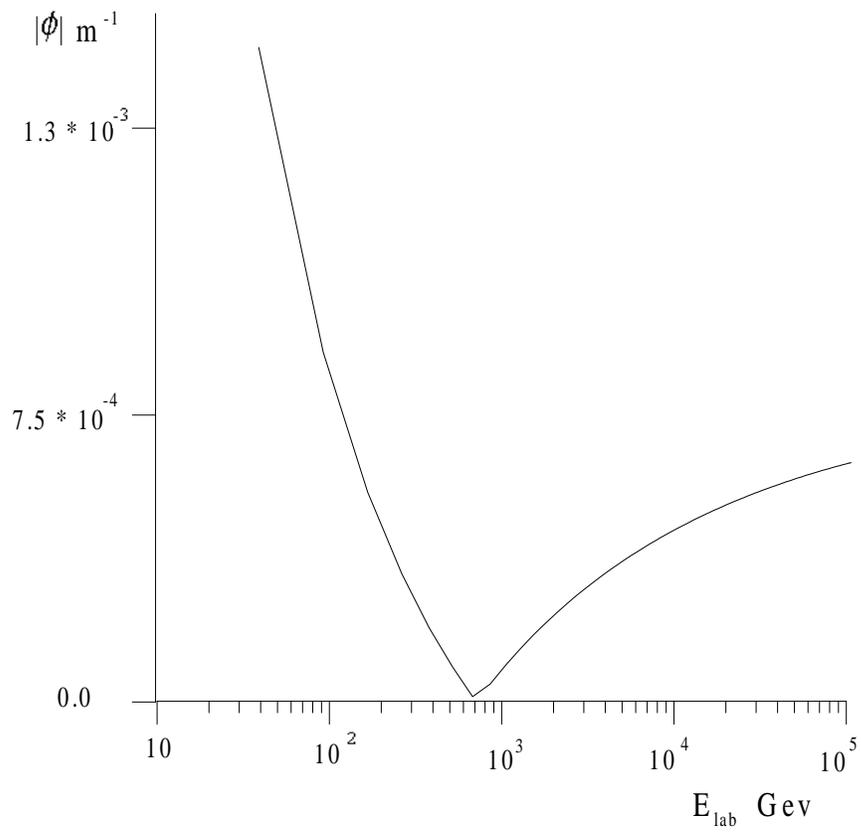}
\caption{
Absolute value of the rotational angle at high
energies.
}
\label{abs}
\end{figure}

\end{document}